\documentclass[twocolumn,prd,superscriptaddress,floatfix,%
preprintnumbers,nofootinbib]{revtex4}
\usepackage{epsfig}

\begin{document}

\title{How robust are inflation model and dark matter constraints from cosmological data?}

\author{Jan Hamann}
\affiliation{Deutsches Elektronen-Synchrotron DESY, Notkestrasse 85,
22607 Hamburg, Germany}

\author{Steen Hannestad}
\affiliation{Department of Physics and Astronomy, University
of Aarhus, Ny Munkegade, DK-8000 Aarhus C, Denmark}

\author{Martin S.~Sloth}
\affiliation{Department of Physics and Astronomy, University
of Aarhus, Ny Munkegade, DK-8000 Aarhus C, Denmark}

\author{Yvonne~Y.~Y.~Wong}
\affiliation{Max-Planck-Institut f\"ur Physik
 (Werner-Heisenberg-Institut), F\"ohringer Ring 6, 80805 M\"unchen,
 Germany}

\date{\today}

\preprint{DESY 06-205, MPP-2006-151}

\begin{abstract}
High-precision data from observation of the cosmic microwave
background and the large scale structure of the universe provide
very tight constraints on the effective parameters that describe
cosmological inflation.  Indeed, within a constrained class of
$\Lambda$CDM models, the simple $\lambda \phi^4$ chaotic inflation
model already appears to be ruled out by cosmological data. In
this paper, we compute constraints on inflationary parameters
within a more general framework that includes other physically
motivated parameters such as a nonzero neutrino mass. We find that
a strong degeneracy between
the tensor-to-scalar ratio $r$ and the neutrino mass prevents
$\lambda \phi^4$ from being excluded by present data.  Reversing
the argument, if $\lambda \phi^4$ is the correct model of
inflation, it predicts a sum of neutrino masses at $0.3\to0.5$ eV,
a range compatible with present experimental limits and within the
reach of the next generation of neutrino mass measurements. We
also discuss the associated constraints on the dark matter
density, the dark energy equation of state, and spatial curvature,
and show that the allowed regions are significantly altered.
Importantly, we find an allowed range of
$0.094 < \Omega_c h^2 < 0.136$ for the dark matter density, a factor
of two larger than that reported in previous studies.  This expanded parameter
space may have  implications for constraints on SUSY dark matter models.
\end{abstract}


\maketitle

\section{Introduction}                           \label{sec:introduction}

The past few years have seen a dramatic increase in the precision
of cosmological data, ranging from measurements of the cosmic microwave
background (CMB) anisotropies by the Wilkinson Microwave Anisotropy
Probe (WMAP) satellite \cite{Spergel:2006hy,Hinshaw:2006ia,Page:2006hz} and
the large scale structure (LSS) of the universe by the
Sloan Digital Sky Survey (SDSS) \cite{Tegmark:2006az,Percival:2006gt},
to the observation of
distant type Ia supernovae (SNIa) \cite{Astier:2005qq}.
All these measurements point to the so-called
concordance model of cosmology, wherein the physical parameters are the baryon
density $\Omega_b$, the matter density $\Omega_m$, the
dark energy density $\Omega_\Lambda$, and the present
Hubble expansion rate $H_0$. The model geometry is flat so that
$\Omega_\Lambda = 1-\Omega_m$, and the initial perturbations are
assumed to be adiabatic and Gaussian,
with a power law spectrum described by a spectral index $n_s$ and
an amplitude $A_s$.  Together with the optical depth parameter $\tau$,
this six-parameter ``vanilla'' model provides a good fit to all
observational data to date.

A common assumption in cosmological parameter estimation is that
one can always improve a fit marginally by including extra free
parameters. This assumption has led to the adoption by many
authors of the Occam's razor approach, in which an extra parameter
is retained only if by its inclusion the goodness-of-fit of the
model is substantially improved. Indeed,
the success of the vanilla model is rooted in the fact that,
given the current data,  no addition of a single
extra parameter produces a $\chi^2$ value that is significantly
lower.

However, there are many more physically well motivated parameters
beyond the vanilla model.
Indeed, some of these, such as a nonzero neutrino mass,
are known to be present.
In such cases, a blind enforcement of Occam's rule
can lead to significant underestimation of parameter
errors, as well as bias in the parameter estimates.
One well-known example is the
interplay between the dark energy equation of state and the
neutrino mass
\cite{Hannestad:2005gj,DeLaMacorra:2006tu,Zunckel:2006mt}. When
the dark energy equation of state is allowed to vary, the neutrino
mass bound is relaxed by almost a factor of three if only CMB and
LSS power spectrum information is used.
Conversely, by imposing a prior on the neutrino masses according to
the Heidelberg--Moscow claims \cite{Klapdor-Kleingrothaus:2001ke,%
Klapdor-Kleingrothaus:2004wj,VolkerKlapdor-Kleingrothaus:2005qv},
reference \cite{DeLaMacorra:2006tu} finds that a cosmological
constant is ruled out at more than $95 \ \%$ C.L.\ by CMB+LSS+SNIa
data.

One could argue that parameter estimation coupled with Occam's rule
is a ``bottom-up'' approach, for which a full Bayesian analysis complete
with Bayes factor calculations may also be
appropriate \cite{Mukherjee:2005wg,Parkinson:2006ku}. However, if one's
aim is to {\it exclude} specific models, then a more conservative
approach that takes into account possible degeneracies between the
``standard''
and the ``new'' parameters is warranted.
Such a ``top-down'' approach does not necessarily imply a decrease in
the predictability of the model.
In fact, we will show that given the present cosmological data,
a nonvanishing neutrino mass could be viewed as
a prediction of the $\lambda\phi^4$ inflationary model. We
argue that when constraining or excluding specific
theoretical models, one should in principle allow for uncertainties in
all physically well-motivated parameters, even if they have
{\it a priori} no direct link to the models concerned. If, for
instance, it turns out later that the universe is indeed composed
of a nonvanishing neutrino fraction, it would be counterproductive
to have already discarded a model of inflation that predicts this
outcome.

In the present work, we investigate in this spirit how parameter constraints
change when the parameter estimation analysis is performed within a
much more general model framework.
In principle there are some twenty or more parameters that
could influence cosmology, although the precision of the
present data is not yet sufficient to probe some of them
(e.g., the primordial helium fraction and the
effective sound speed of dark energy).
Here, we focus on a 11-parameter model outlined below.

\subsection{The model}

We test a general 11-parameter model space consisting of
\begin{equation}
\Theta =
(\omega_c,\omega_b,f_\nu,\Omega_k,w,H_0,n_s,r,\alpha_s,\tau,A_s).
\end{equation}
The vanilla model is defined by $f_\nu=\Omega_k=r=\alpha_s=0$,
and $w=-1$.  In addition, we marginalise over a nuisance parameter $b$
which describes the relative
bias between the observed galaxy power spectrum $P_g(k)$ and the
underlying dark matter spectrum $P_c(k)$ via $P_c(k)=b^2 P_g(k)$.

Three different parameter sets will be considered in this work:

\begin{itemize}
\item Set A: All 11 parameters.
\item Set B: A 10-parameter set with $\Omega_k=0$.  This parameter set covers
all standard inflationary models
\item Set C: A 9-parameter set with $\Omega_k = \alpha_s = 0$.  This reduced set corresponds
to the large subset of the zoo of inflationary models that predict negligible running,
including large field chaotic inflation models \cite{linde}.
\end{itemize}

\subsubsection{\it Matter content}

We assume the matter content to be specified by the following
parameters: the curvature $\Omega_k = 1-\Omega_m-\Omega_\Lambda$,
the physical dark matter density $\omega_c=\Omega_{\rm c} h^2$,
the baryon density $\omega_b=\Omega_b h^2$, the neutrino fraction
$f_\nu = \Omega_\nu/\Omega_c$,  and the dark energy equation of state parameter
$w$.

Other parameters not included here, but
which could have an observable effect, include a time-dependent
dark energy equation of state, nonstandard interactions in any of
the dark sectors (cold dark matter, neutrinos, or dark energy), etc. We mention
this as a caution that while our parameter space is much larger
than that normally used in parameter estimation analyses, it is not
necessarily complete.

\begin{figure}[t]
\begin{center}
\epsfig{file=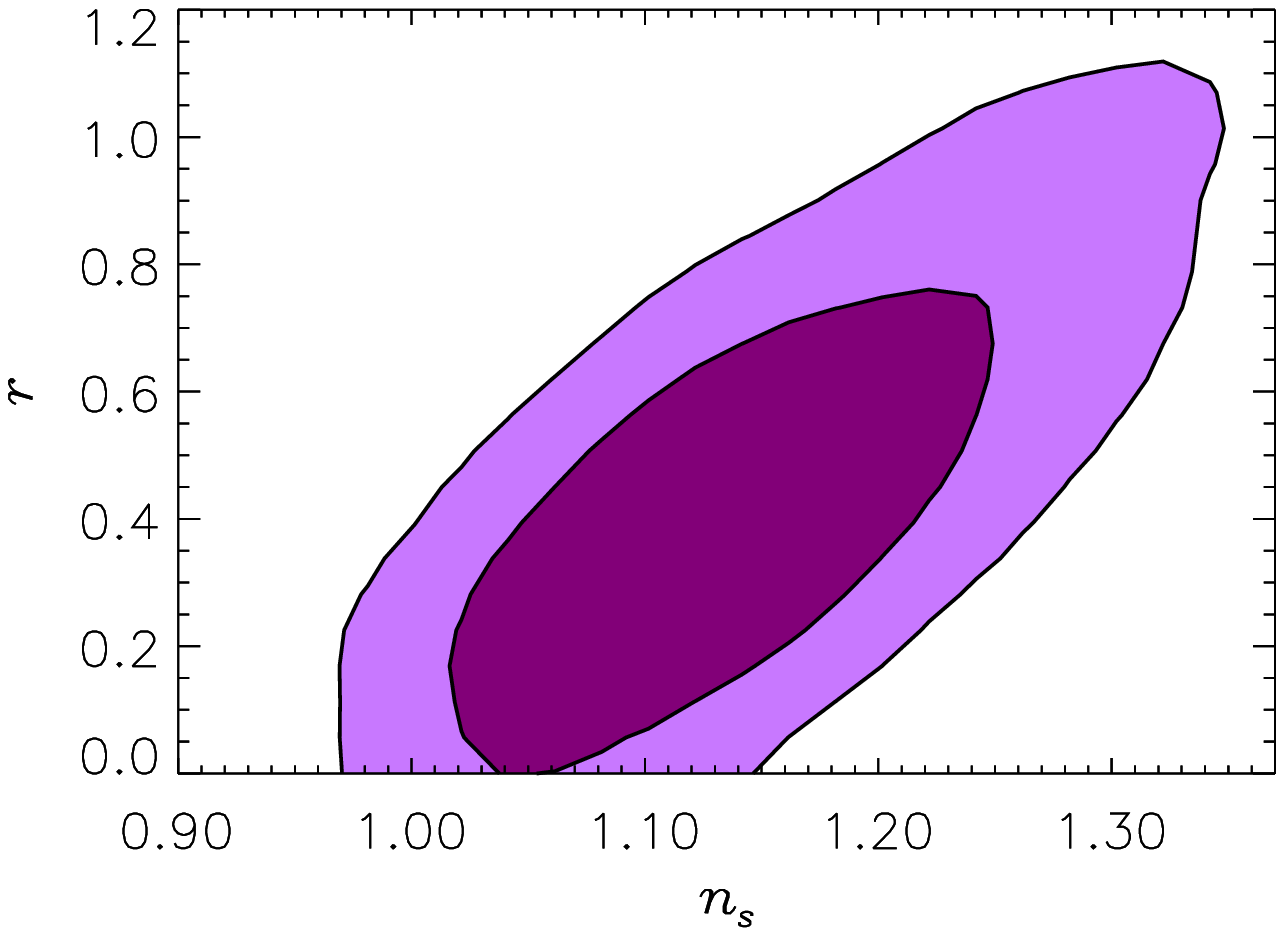,width=8.5cm}\vspace*{-0.5cm}\\
\epsfig{file=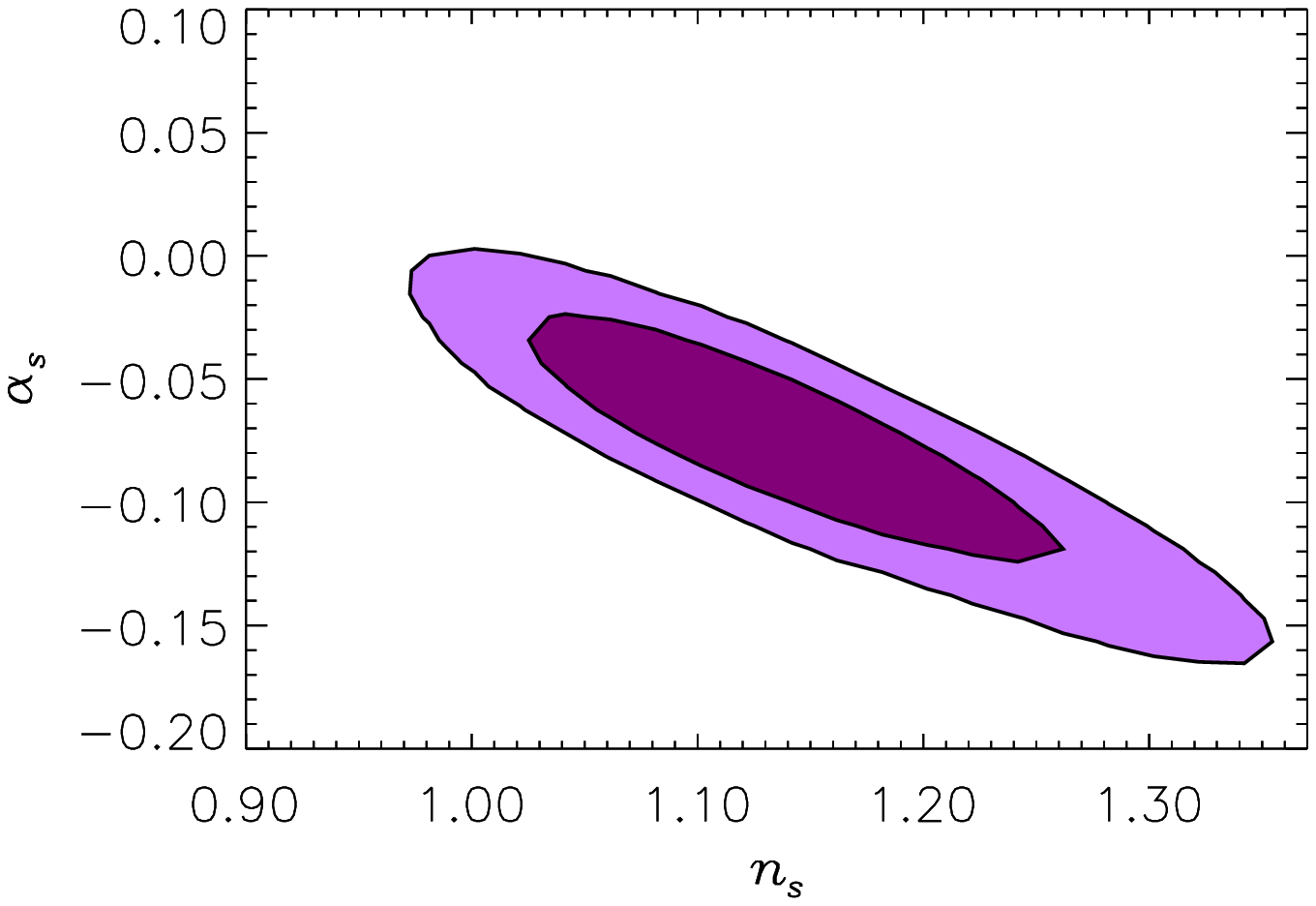,width=8.5cm}\vspace*{-0.5cm}\\
\epsfig{file=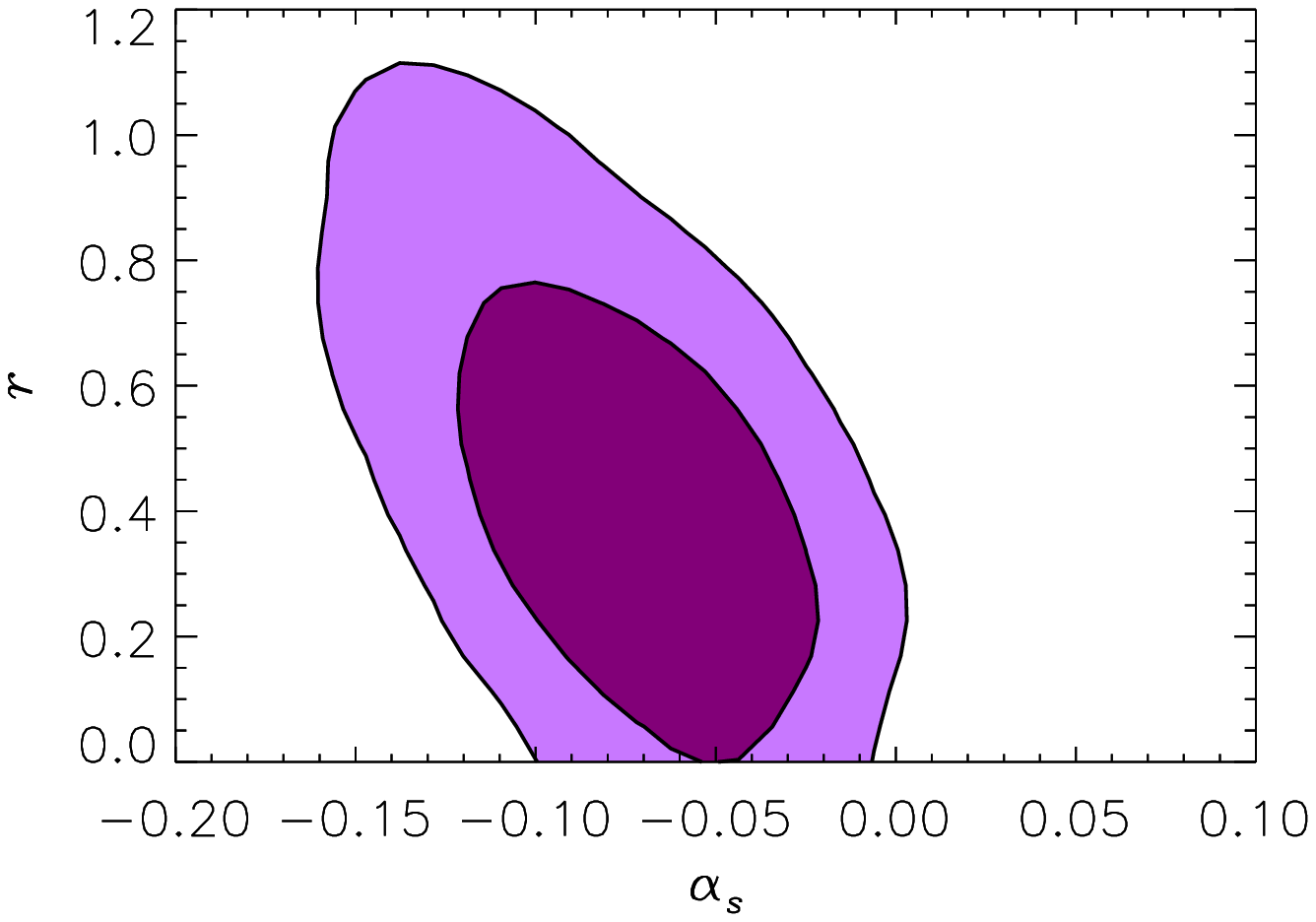,width=8.5cm}
\end{center}
\caption{Two-dimensional $68\ \%$ and $95 \ \%$ C.L.\ contours for the
inflationary parameters $n_s$, $r$,
and $\alpha_s$, using the full data set and parameter set B, and
marginalised over
the other $(10-2)$ parameters.}\label{fig:large}
\end{figure}

\subsubsection{Initial conditions}

The initial conditions for structure formation are assumed to be
set by inflation, characterised by scalar and tensor fluctuations
with amplitude $A_s$ and $A_t = r A_s$ respectively. Each
component is specified by a spectral index $n_s$ or $n_t$,
and the inflationary consistency relation requires that $n_t \sim
-r/8$. However, the precision of current data is not yet at a
level where a violation of the consistency relation can be tested.
This also means that while the running parameter $\alpha_s$ should
be included for the scalar spectrum, the inclusion of its tensor
counterpart $\alpha_t$ would have no effect. This set of initial
parameters encompasses all standard inflationary models, but not
models with features from potential steps, particle production, etc.\
during inflation. We define $\alpha_s$ at the pivot scale
$k=0.002$ Mpc$^{-1}$, in concordance with most recent analyses.

Note that an alternative approach would be to perform the analysis
directly in terms of the slow-roll parameters instead of the
observables $n_s$, $\alpha_s$, and $r$
\cite{Leach:2003us,Peiris:2006ug,Peiris:2006sj}. Particularly for
models where $\alpha_s$ is not negligible this can lead to somewhat
different results. However, in models with small $\alpha_s$, such as
chaotic inflation, the results are identical.

\section{Data analysis}                 \label{sec:data}

\begin{figure}[t]
\begin{center}
\epsfig{file=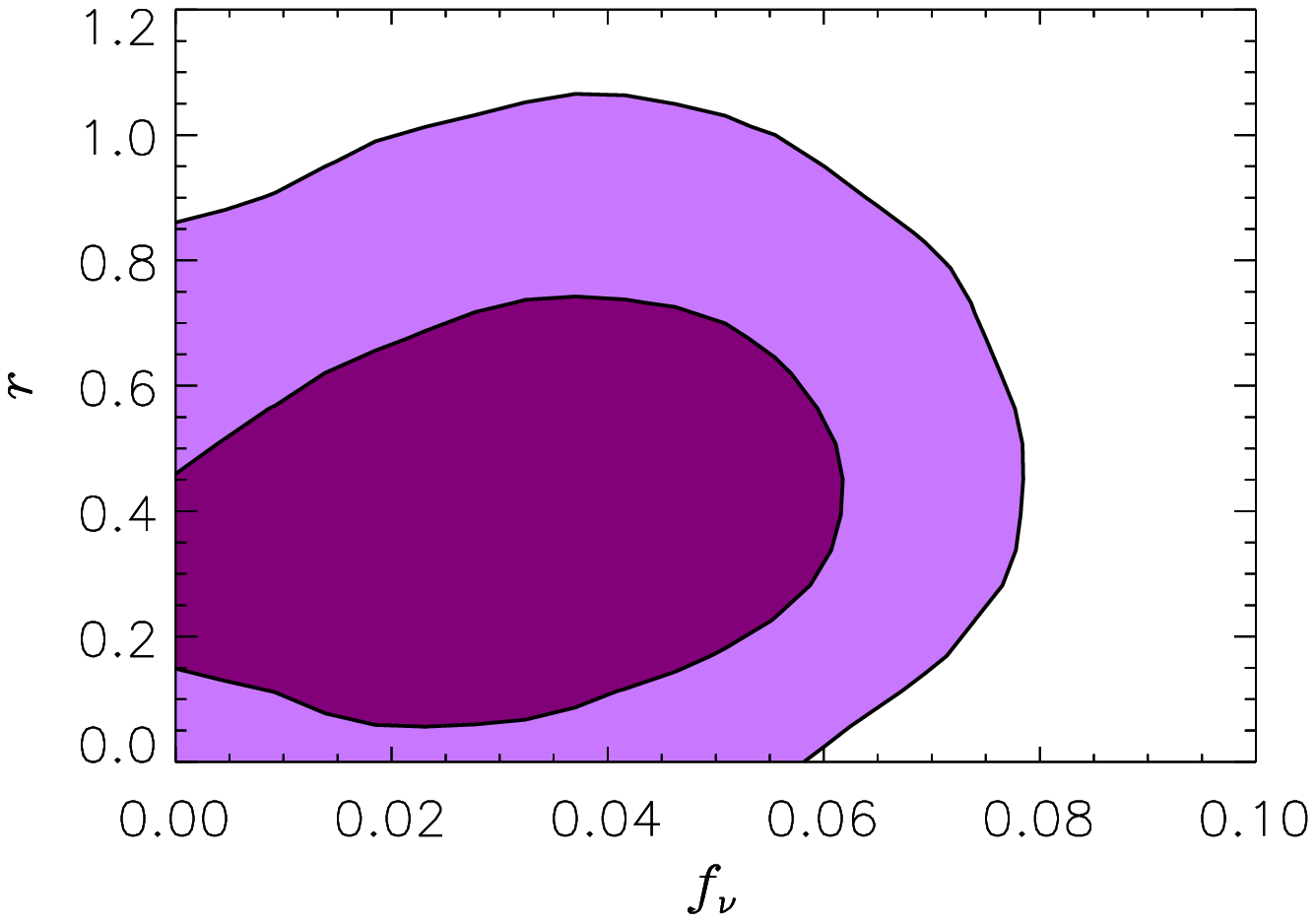,width=8.5cm}\vspace*{-0.5cm}\\
\epsfig{file=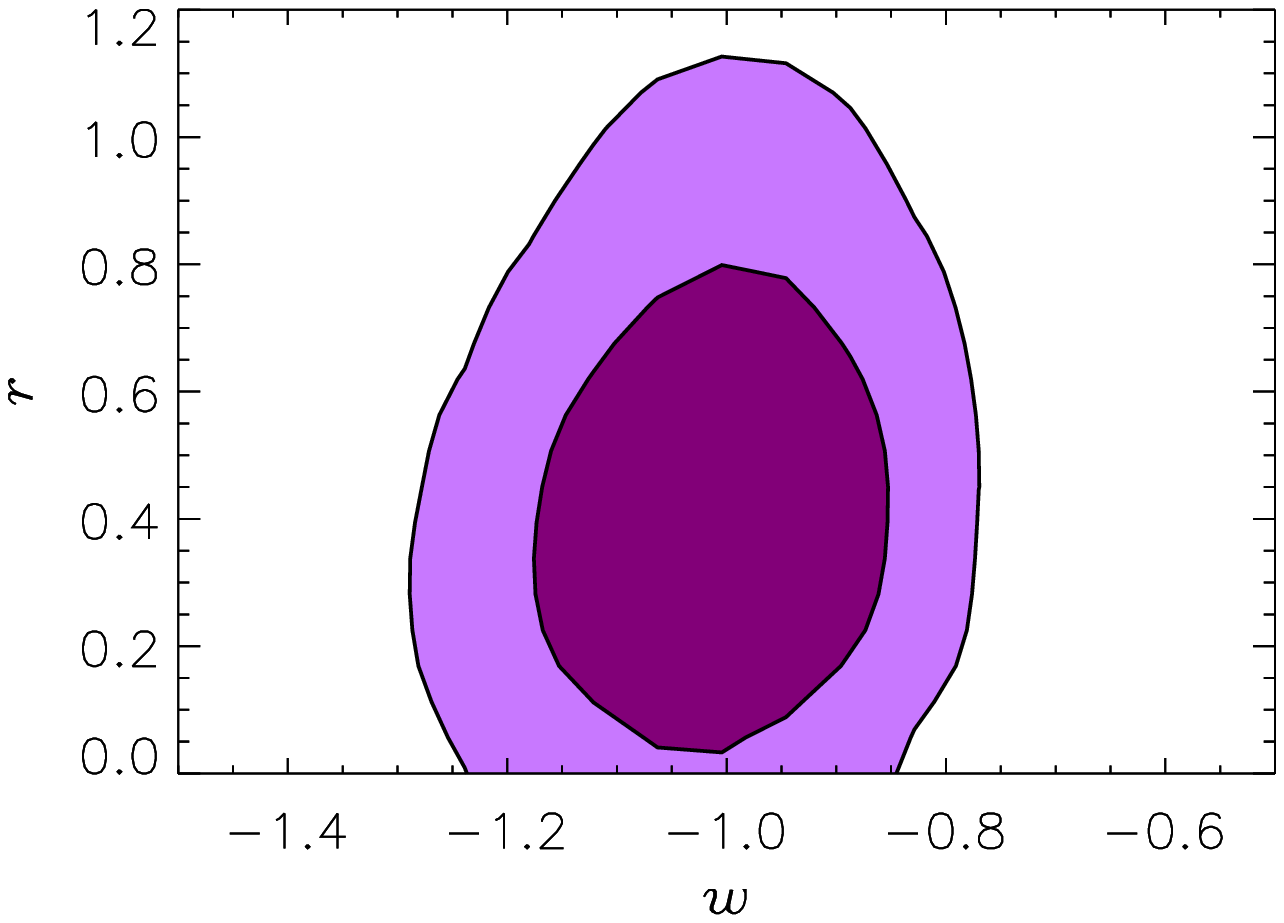,width=8.5cm}
\end{center}
\caption{Degeneracies between $r$ and $f_\nu,w$ for the full data
set and parameter set B, marginalised over $(10-2)$ parameters.}\label{fig:degeneracy}
\end{figure}

{\it Cosmic microwave background (CMB)} \quad
We use CMB data from the WMAP experiment after three years of
observation \cite{Spergel:2006hy,Hinshaw:2006ia,Page:2006hz}. The data
analysis is performed using the likelihood calculation package
provided by the WMAP team on the LAMBDA homepage \cite{lambda}.

{\it Large scale structure (LSS)} \quad
The large scale structure power spectrum of luminous
red galaxies (LRG) has been measured by the Sloan Digital Sky
Survey (SDSS). We use the same analysis technique on this data set as
advocated by the SDSS team \cite{Tegmark:2006az,Percival:2006gt},
with analytic marginalisation over the bias $b$
and the nonlinear correction parameter $Q_{\rm nl}$.

{\it Baryon acoustic oscillations (BAO)} \quad
In addition to the power spectrum data we use the measurement of
baryon acoustic oscillations in the two-point correlation function
\cite{Eisenstein2005}.  The analysis is performed following the
procedure described in \cite{Eisenstein2005,Eisensteinweb}
(see also \cite{Goobar:2006xz}),
 including
analytic marginalisation over the bias $b$, and
nonlinear corrections with the HALOFIT \cite{halofit} package.

{\it Type Ia supernovae (SNIa)} \quad
We use the luminosity distance measurements of
distant type Ia supernovae provided by the Supernova Legacy Survey (SNLS)
\cite{Astier:2005qq}.

{\it Lyman-$\alpha$ forest} \quad We do not include data from the
Lyman-$\alpha$ forest in our analysis. These data were used
in some previous studies that found very strong bounds on various cosmological
parameters~\cite{seljak2006}. However, the strength of these bounds
is due mainly to the fact that the Lyman-$\alpha$ analysis
used in \cite{seljak2006} leads to a much higher
normalisation of the small-scale power spectrum than that obtained from
the WMAP
data.  Other analyses of the same SDSS Lyman-$\alpha$ data find a
lower normalisation, in better agreement with the WMAP
result~\cite{Viel:2005eg,Viel:2005ha,viel2006}. This kind of discrepancy
between different analyses of the same data probably points to
unresolved systematic issues, and for this reason we prefer to discard the
Lyman-$\alpha$ data entirely.

For a large part of our analysis we use two different combinations of data sets,
one consisting of WMAP and SDSS data only, and one which uses
in addition data from SNIa (SNLS) and BAO.
The latter case is sometimes referred to as ``the full data set''.

We perform the data analysis using the publicly available CosmoMC
package \cite{Lewis:2002ah,cosmomc}, modified to include the
BAO likelihood calculations.

\begin{table}[t]
\caption{\label{tab:inflation1}  The $95 \ \%$ C.L.\ allowed ranges for
$n_s$, $r$ and $\alpha_s$ for parameter set B, marginalised over the
other $(10-1)$ parameters.}
\begin{ruledtabular}
\begin{tabular}{lcc}
Parameter & WMAP+SDSS & Full data set \\
\hline
$n_s$ & $0.97 \to 1.35$ & $0.98 \to 1.28$ \\
$r$ & $0 \to 1.05$ & $0 \to 0.81$ \\
$\alpha_s$ & $-0.140 \to -0.005$ & $-0.135 \to -0.004$ \\
\end{tabular}
\end{ruledtabular}
\end{table}

\section{Results}                                \label{sec:results}

\subsection{Inflationary parameters}

Almost all inflationary models predict $\Omega_k$ to be
zero.  This prediction is also supported by our analysis of parameter
set A (see Sec.~\ref{sec:DMandDE}).  Therefore, in this section, we will
work with the reduced 10-parameter set B, in which $\Omega_k$ is already
fixed at zero.
Figure~\ref{fig:large} shows the 2D
likelihood contours
for the  parameters $n_s,r$ and $\alpha_s$ using the full data set
and parameter set B.  These contours are obtained by marginalising over
the other $(10-2)$ parameters not shown in the plot.

Figure~\ref{fig:large} should be compared with, e.g., Figs.~2 and 3 of
Kinney {\it et al.} \cite{Kinney:2006qm}, which use data from
WMAP and SDSS, and a parameter set similar
to our set B but with $f_\nu$ and $w$ fixed at $0$ and $-1$ respectively.
The comparison reveals that the two sets of likelihood contours are roughly similar,
but with one important exception: the allowed range for the tensor-to-scalar ratio
$r$ in our case
is much larger even in the light of additional data!

In order to understand this effect we plot in Fig.~\ref{fig:degeneracy}
the 2D likelihood contours for $r$ and our additional parameters
$f_\nu$ and $w$.  Interestingly, a substantial degeneracy exists between
$r$ and the neutrino fraction $f_\nu$, which in turn allows $r$ to extend
to much higher values. Table \ref{tab:inflation1} displays the 1D $95 \ \%$
C.L.\ allowed ranges for $n_s$, $r$, and $\alpha_s$, assuming parameter set B
and using both WMAP+SDSS only and
the full data set.

\begin{figure}[t]
\begin{center}
\epsfig{file=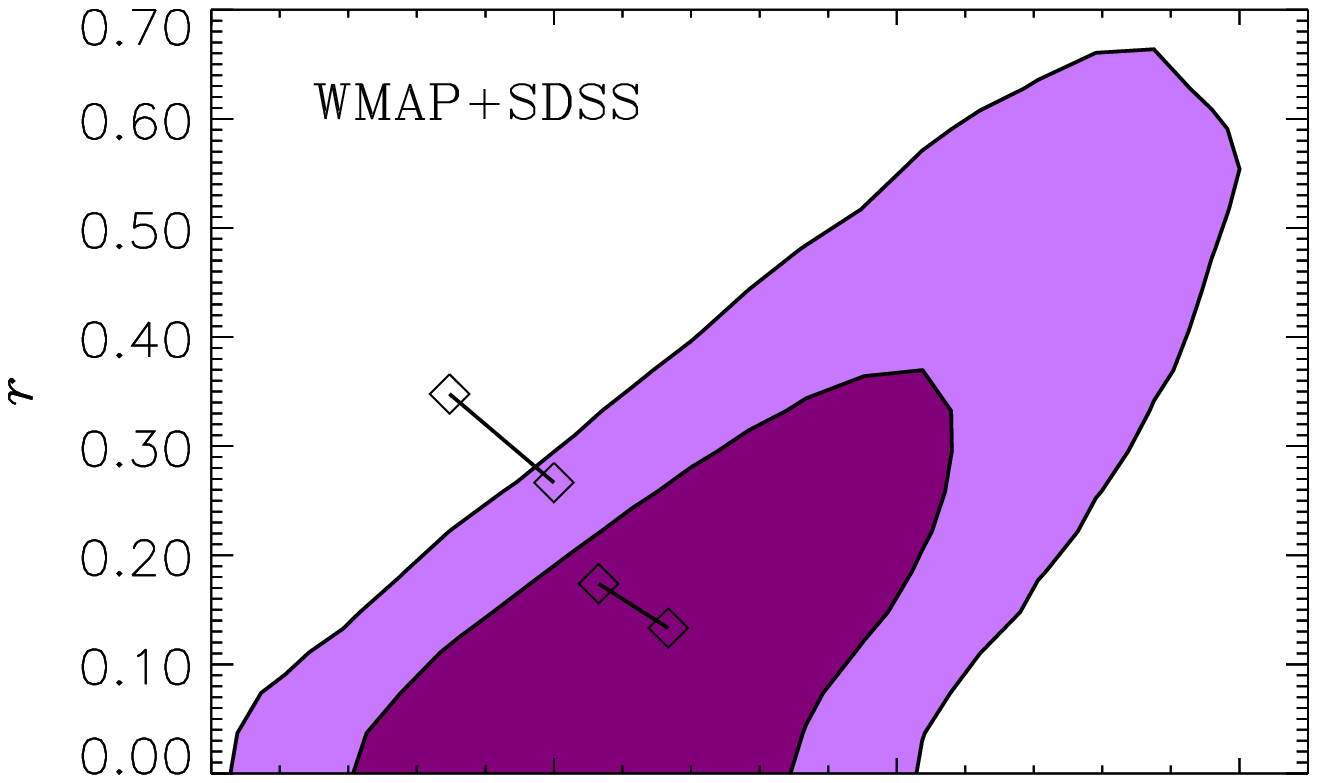,width=8.5cm} \vspace*{-1cm}\\
\epsfig{file=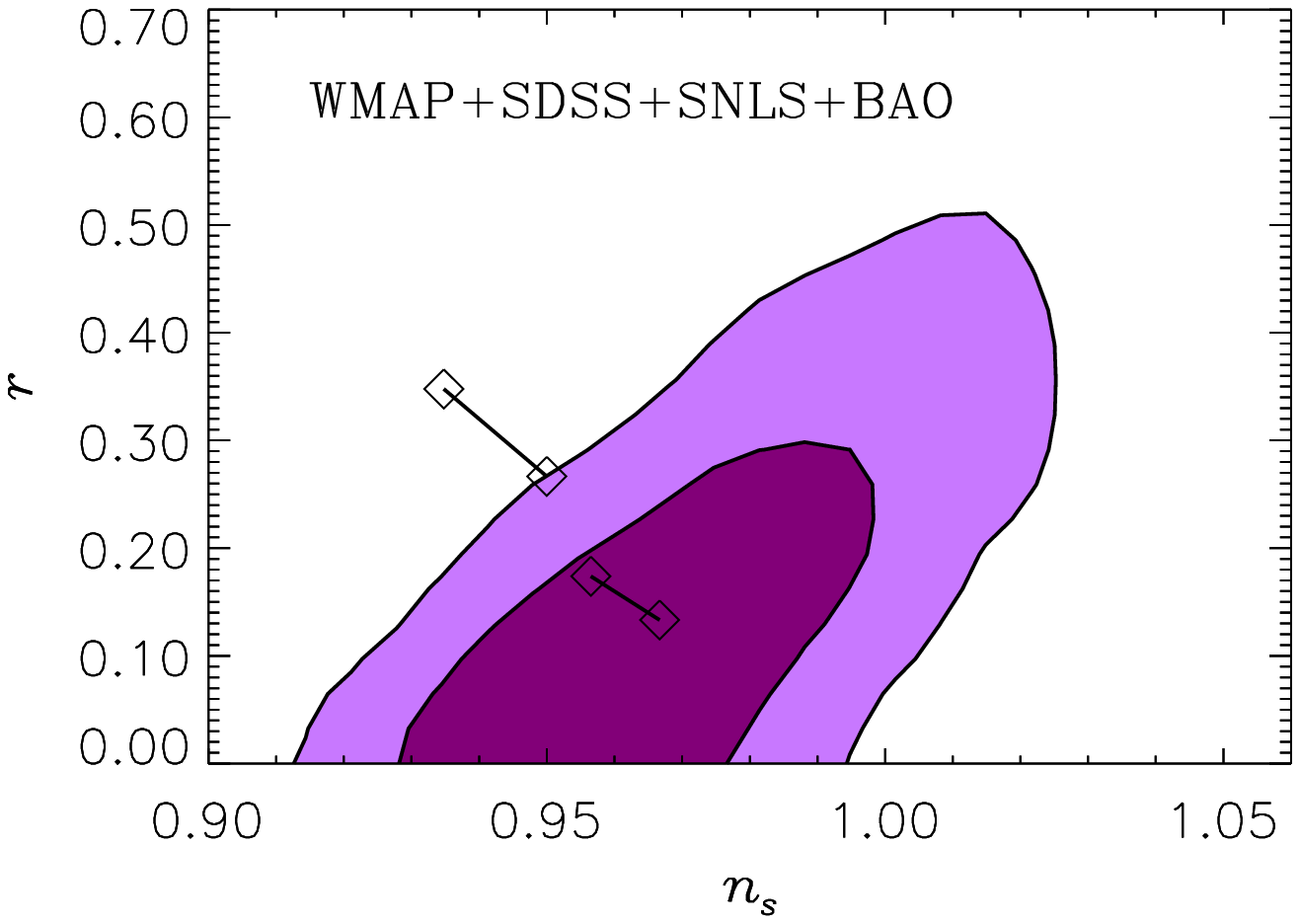,width=8.5cm}
\end{center}
\caption{Two-dimensional $68 \ \%$ and $95 \ \%$ C.L.\ contours for $n_s$ and $r$,
using parameter set C (consistent with predictions of chaotic inflation), and
marginalised
over $(9-2)$ parameters.
The upper panel uses WMAP+SDSS data and the lower the full data
set.
The two short black/solid lines with boxes at the ends
correspond to  predictions of $\lambda \phi^4$ (top left) and
$m^2 \phi^2$ models of inflation,
with 46 to 60 $e$-foldings (left to right).
}\label{fig:nr}
\end{figure}

\subsection{Chaotic inflation}

Single field inflation models with polynomial potentials generally
predict negligible running.
These models are thus represented by our 9-parameter set C in which
$\alpha_s=0$.
The corresponding 2D likelihood contours for $n_s$ and $r$,
marginalised over the other $(9-2)$ parameters,
are shown in Fig.~\ref{fig:nr}.

\begin{table*}
\caption{\label{tab:mass}The 1D marginalised $95 \ \%$ C.L.\ allowed ranges for $n_s$
and $r$ for parameter set C and its subsets.}
\begin{ruledtabular}
\begin{tabular}{lccc}
Parameter set & Data set & $n_s$ & $r$ \\
\hline
C & WMAP+SDSS & $0.927 \to 1.038$ & $0 \to 0.51$ \\
C & WMAP+SDSS+SNLS+BAO & $0.932 \to 1.018$ & $0 \to 0.41$ \\
C & WMAP+SDSSlin+SNLS+BAO & $0.931 \to 1.025$ & $0 \to 0.47$ \\
C, $w$ fixed & WMAP+SDSS+SNLS+BAO & $0.933 \to 1.019$ & $0 \to 0.40$ \\
C, $w,f_\nu$ fixed & WMAP+SDSS & $0.931 \to 1.011$ & $0 \to 0.31$ \\
C, $w,f_\nu$ fixed & WMAP+SDSS+SNLS+BAO & $0.931 \to 1.010$ & $0 \to 0.30$ \\
\end{tabular}
\end{ruledtabular}
\end{table*}

Figure~\ref{fig:nr} should be compared with Fig.~4 of Kinney {\it et
al.} \cite{Kinney:2006qm}, with Fig.~14 of Spergel {\it et al.}
\cite{Spergel:2006hy}, and with Fig.~19 of  Tegmark {\it et al.}
\cite{Tegmark:2006az} (see also \cite{Martin:2006rs}).
In all cases our WMAP+SDSS contours encompass
a markedly larger region. In particular, even with the inclusion of
SNIa and BAO data, we find that the simplest $\lambda \phi^4$ model
is still allowed by data, contrary to the conclusions of
\cite{Kinney:2006qm,Spergel:2006hy,Tegmark:2006az,Martin:2006rs}.
We note  that the endpoints of the model lines
in Fig.~\ref{fig:nr} correspond to 46 and 60 e-foldings respectively.%
\footnote{When taking into account one-loop effects in the
chaotic inflationary scenario, the model lines in this plot may actually be
smeared at the percent level \cite{Sloth:2006az}.}
Interestingly, $\lambda \phi^4$ is compatible with data only if the number
of $e$-foldings is relatively large, or equivalently, if the
reheating temperature is high \cite{dodelson,Liddle:2003as}.

\begin{figure}[b]
\begin{center}
\epsfig{file=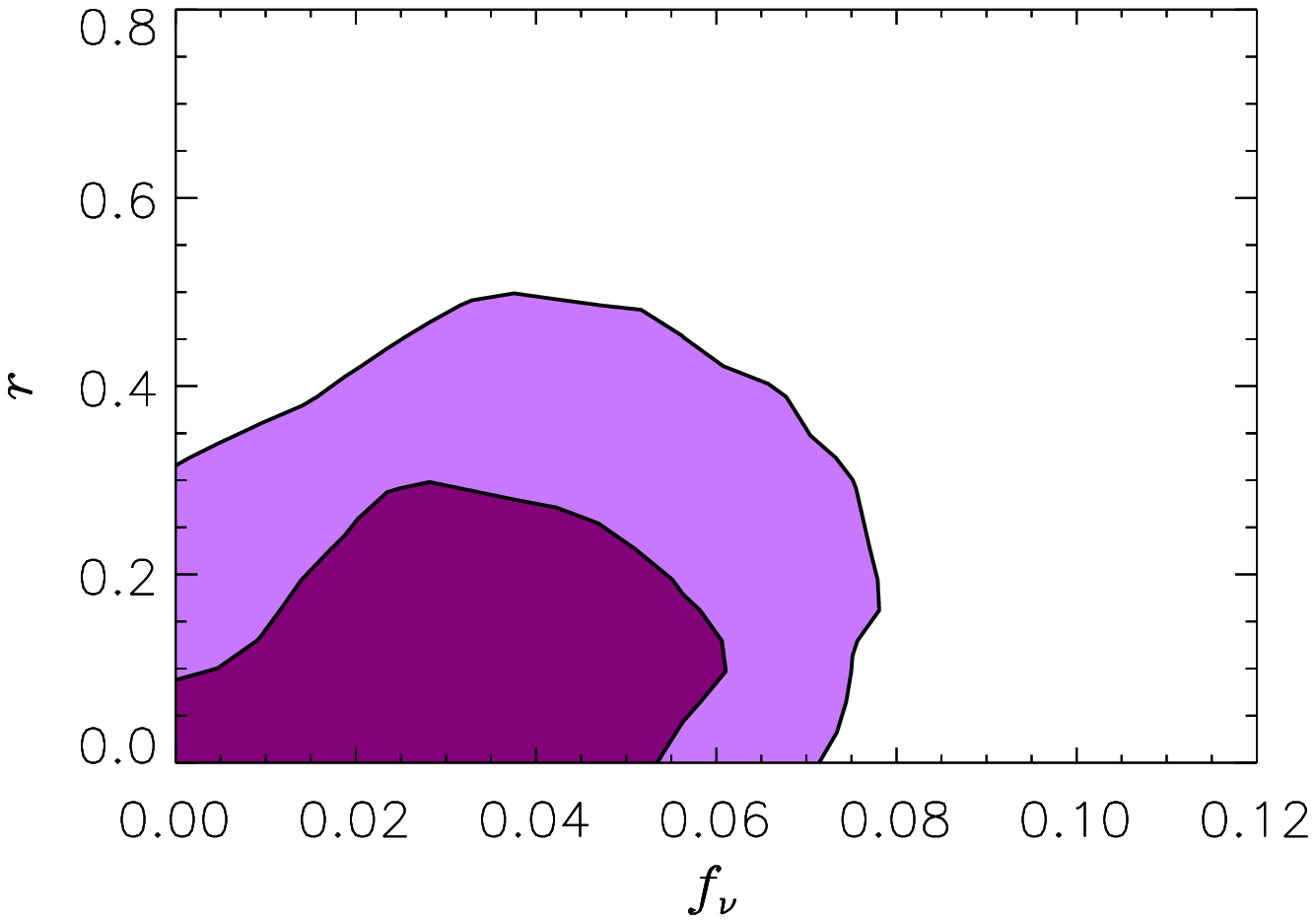,width=8.5cm} \vspace*{-0.5cm}\\
\epsfig{file=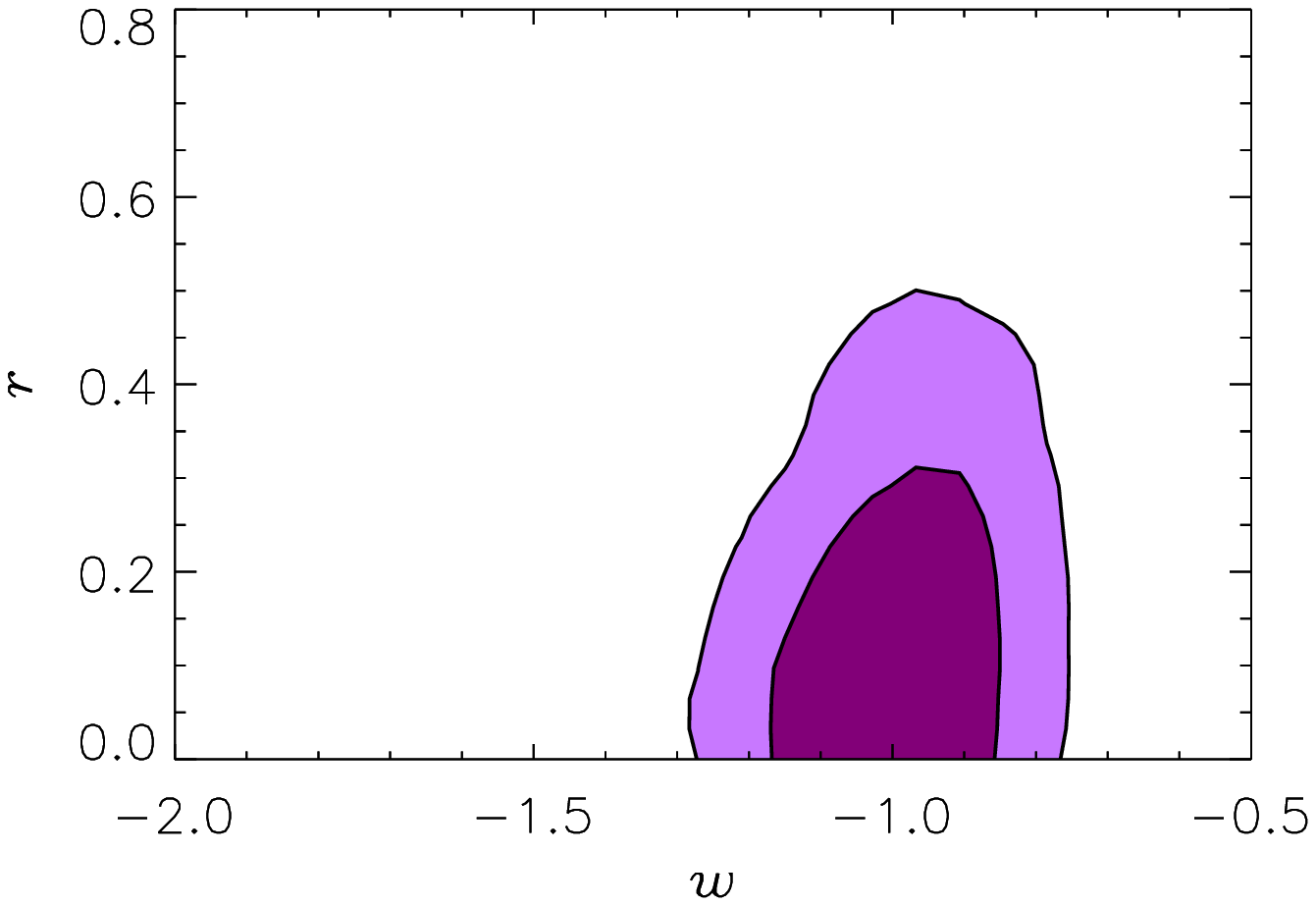,width=8.5cm}
\end{center}
\caption{Degeneracies between between $r$ and $f_\nu,w$ for
the full data set and parameter set C,
marginalised over $(9-2)$ parameters.}\label{fig:deg2}
\end{figure}

Again, the explanation for our enlarged $n_s,r$ allowed region
lies in our expanded model parameter space. In
Fig.~\ref{fig:deg2}, we see that the degeneracy
between $r$ and $f_\nu$
encountered earlier in parameter set B is present also in parameter set C,
albeit to a smaller extent.
If a neutrino fraction of $0.03 \to 0.05$ is allowed
(corresponding roughly to $\sum m_\nu \sim 0.3 \to 0.5$ eV),
new parameter space opens up for $n_s$ and $r$. We note in passing
that the converse is not true. Allowing $r$ to run does not change
the upper bound on the neutrino mass significantly. Interestingly,
this $f_\nu,r$ degeneracy also means that $\lambda \phi^4$ in its simplest
form predicts quasi-degenerate neutrino masses with a sum in
the $0.3 \to 0.5$ eV range.  This range is compatible with present
laboratory limits from tritium beta decay experiments, $m_\nu  < 2.2 \ {\rm eV}$
\cite{Lobashev:2003kt,Kraus:2004zw}, as well as
the claimed detection of neutrinoless double beta decay, and hence
detection of the effective electron neutrino mass $m_{ee} =
\left| \sum_j U^2_{ej} m_{\nu_j} \right|$ at $0.1 \to 0.9$ eV ,
by the Heidelberg--Moscow experiment
\cite{Klapdor-Kleingrothaus:2001ke,Klapdor-Kleingrothaus:2004wj,%
VolkerKlapdor-Kleingrothaus:2005qv}.
The upcoming tritium beta decay experiment KATRIN
will also probe neutrino masses to a comparable level of
precision \cite{katrin}.

\begin{figure}[t]
\begin{center}
\epsfig{file=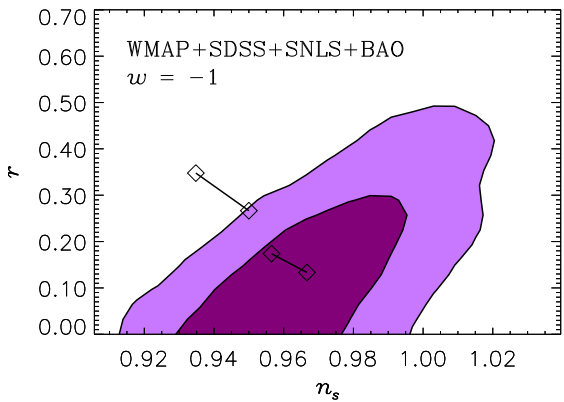,width=8.5cm}
\end{center}
\caption{Two-dimensional marginalised constraints on $n_s$ and $r$
for the parameter set C, but with the restriction $w=-1$.}\label{fig:w}
\end{figure}

Also of interest is the case of a fixed dark energy equation of state
$w$.  Figure~\ref{fig:w} shows the equivalent of the lower panel of
Fig.~\ref{fig:nr} (full data set and parameter set C),
but with the additional restriction $w=-1$. Clearly,
there is very little difference between Figs.~\ref{fig:nr} and \ref{fig:w},
since the combination of SNIa and BAO data effectively fixes
$w$ to $-1$ in the former case, as shown in Fig.~\ref{fig:deg2}.

\begin{figure}[b]
\begin{center}
\epsfig{file=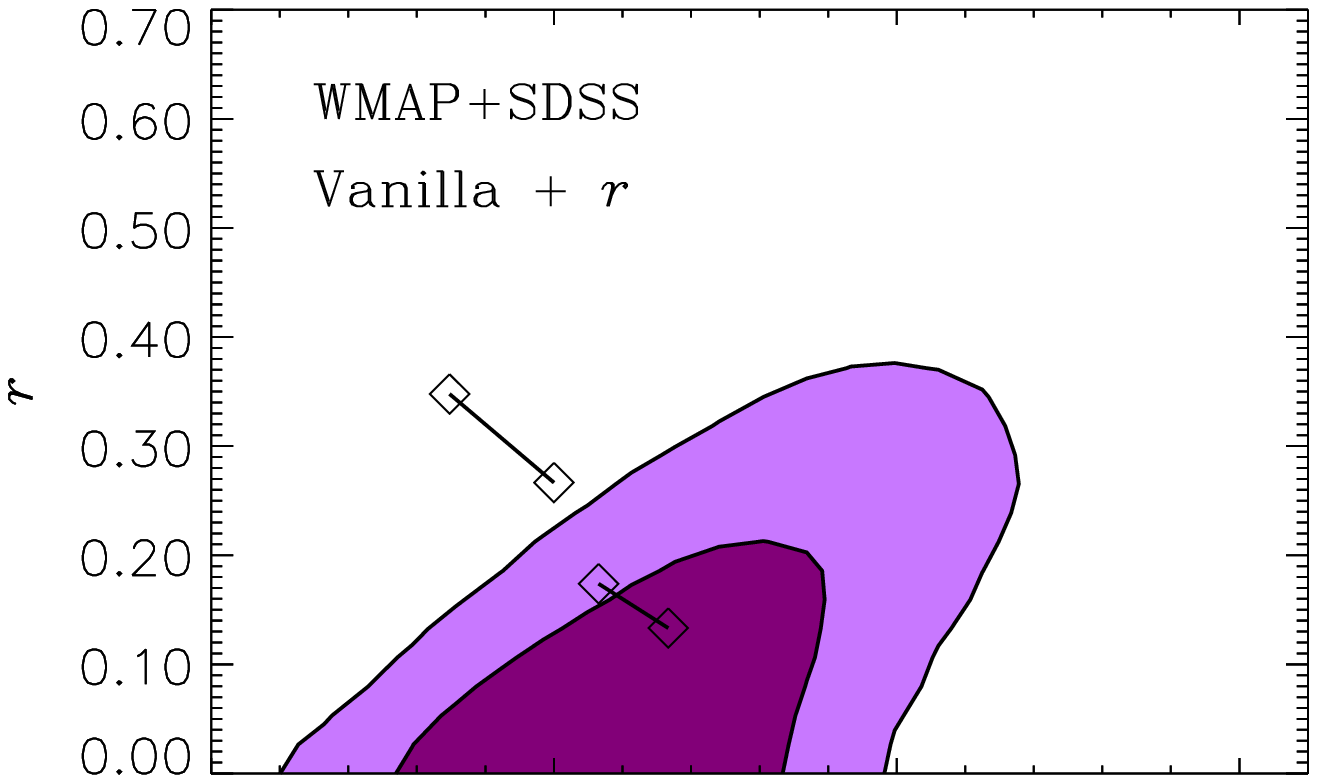,width=8.5cm} \vspace*{-1cm}\\
\epsfig{file=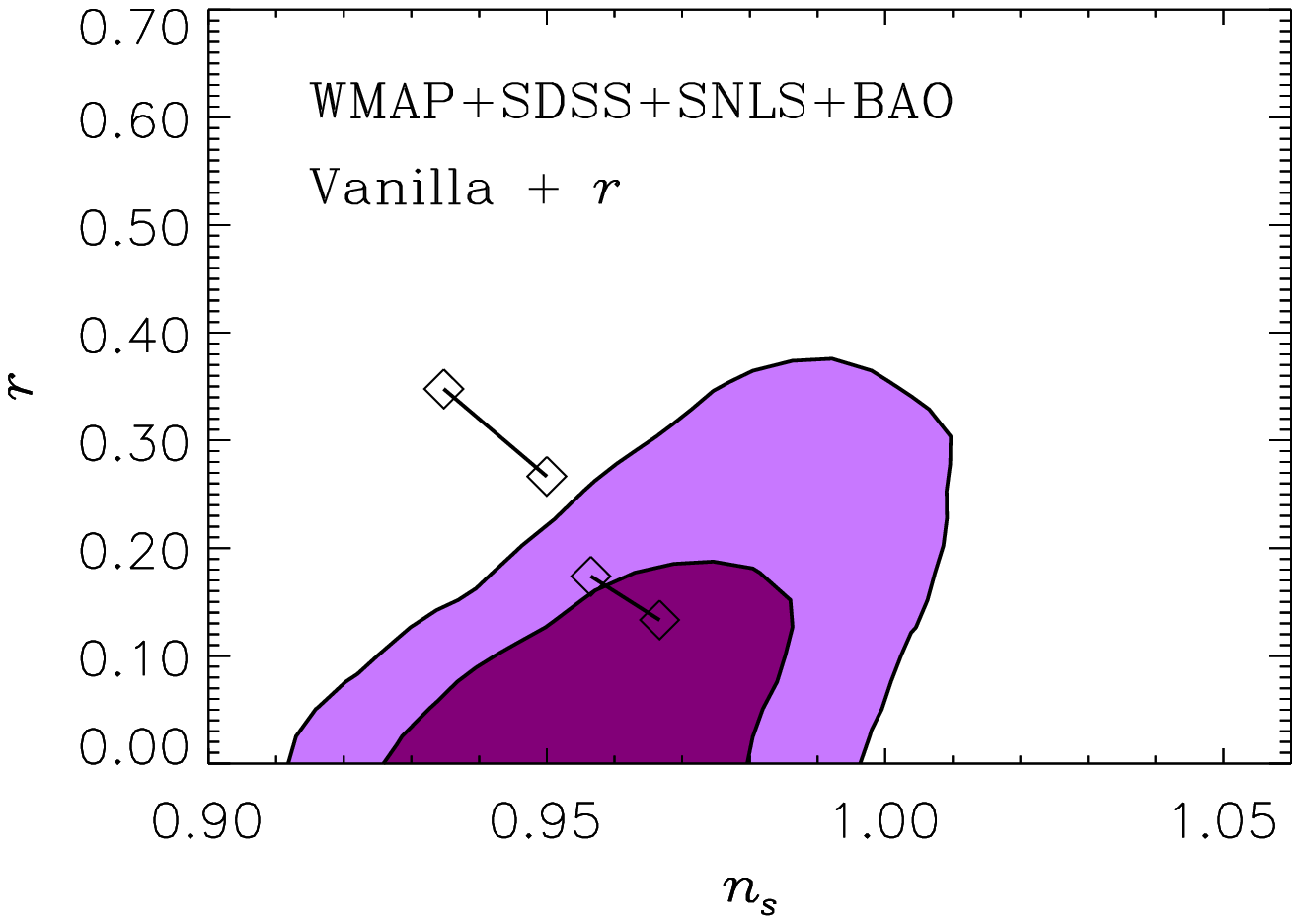,width=8.5cm}
\end{center}
\caption{Two-dimensional marginalised contours for the vanilla+$r$
parameter space, for both WMAP+SDSS only (upper panel) and the full
data set (lower panel).}\label{fig:comp}
\end{figure}

As a consistency check we present in Fig.~\ref{fig:comp} also the 2D
constraints on $n_s,r$ for the vanilla model with one extra
parameter $r$, i.e., the same model analysed in
\cite{Kinney:2006qm,Spergel:2006hy,Tegmark:2006az}. The general
shapes of the contours in this figure are almost identical to those
in Fig.~19 in Tegmark {\it et al.} \cite{Tegmark:2006az} which uses
the same data sets.   In addition, we find a 1D $95 \ \%$ C.L.\
upper bound of $r<0.31$, while Tegmark {\it et al.} report an almost
identical $r<0.33$. Kinney {\it et al.} also found $r<0.31$ for the
same vanilla+$r$ model \cite{Kinney:2006qm}, but from a combination
of WMAP and the SDSS main galaxy samples (as opposed to SDSS LRG
used in this work and in \cite{Tegmark:2006az}). For comparison
\cite{finelli} found $r<0.26$ for an analysis of WMAP and 2dF.

Using additional data from the Lyman-$\alpha$ forest, Seljak {\it
et al.} \cite{seljak2006} derived an even stronger upper bound,
$r<0.22$, for the same model space. The reason for the improvement
is a degeneracy between $r$ and $\sigma_8$, such that a higher value of $r$
leads to a smaller preferred value of $\sigma_8$. Since the
Lyman-$\alpha$ data used in \cite{seljak2006} prefer a high value
of $\sigma_8$, a small $r$ value is correspondingly favoured.  In fact,
from a parameter fitting point of view, a negative $r$ would be even
better.  All these conspire to give a much stronger upper bound on $r$.
However, as noted in Sec.~\ref{sec:data}, this phenomenon likely points
to a systematic uncertainty in the Lyman-$\alpha$ normalisation, rather
than a genuinely strong constraint on $r$.

Finally, we stress again that the difference between the
allowed $n_s,r$ regions in Figs.~\ref{fig:nr} and
\ref{fig:comp} lies in a degeneracy between $r$ and the neutrino
fraction $f_\nu$. It should also be noted that
the addition of SNIa and BAO data has very little impact
on the vanilla+$r$ model, because no strong parameter degeneracies
are present in the WMAP+SDSS data.
With SNIa and BAO included we find a 1D $95 \ \%$ C.L.\ bound of
$r<0.30$, instead of $0.31$ for WMAP+SDSS alone.

\begin{figure}[t]
\begin{center}
\epsfig{file=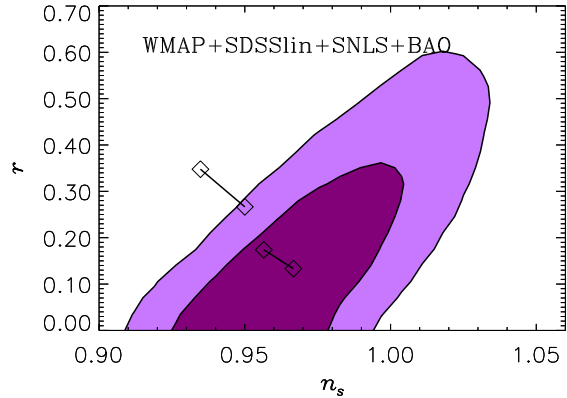,width=8.5cm}
\end{center}
\caption{Two-dimensional marginalised constraints on $n_s$ and $r$ for
parameter set C. Only the linear part of the SDSS power
spectrum has been used.}\label{fig:linear}
\end{figure}

\subsection{The effect of non-linearity}

So far we have used exactly the same analysis technique as the
SDSS team when treating the LRG data. However, beyond a wavenumber
of approximately $k \sim 0.06 \to 0.07 \ h \ {\rm  Mpc}^{-1}$,
nonlinear effects begin to dominate the matter spectrum
(see, for instance, Fig.~9 of \cite{Tegmark:2006az}). To test
whether or not our results are subject to these effects,
we perform the same analysis as in Fig.~\ref{fig:nr}, but retain
data only up to $k \sim 0.06 \ h \ {\rm Mpc}^{-1}$ (band 11).
We call this reduced data set SDSSlin, and the result is shown
in Fig.~\ref{fig:linear}. Using only the linear part of the
power spectrum data has no bearing on our conclusions. In fact,
the 2D allowed region in $n_s,r$  for parameter set C
is only affected in the region where $n_s>1$.
The SDSS data probes $n_s$ more precisely when all data
points are included, and this in turn leads to a truncation of the
allowed region at high $n_s$.

Table \ref{tab:mass} summarises the 1D marginalised
constraints on $n_s$ and $r$ for parameter set C and its subsets.

\begin{figure}[b]
\begin{center}
\epsfig{file=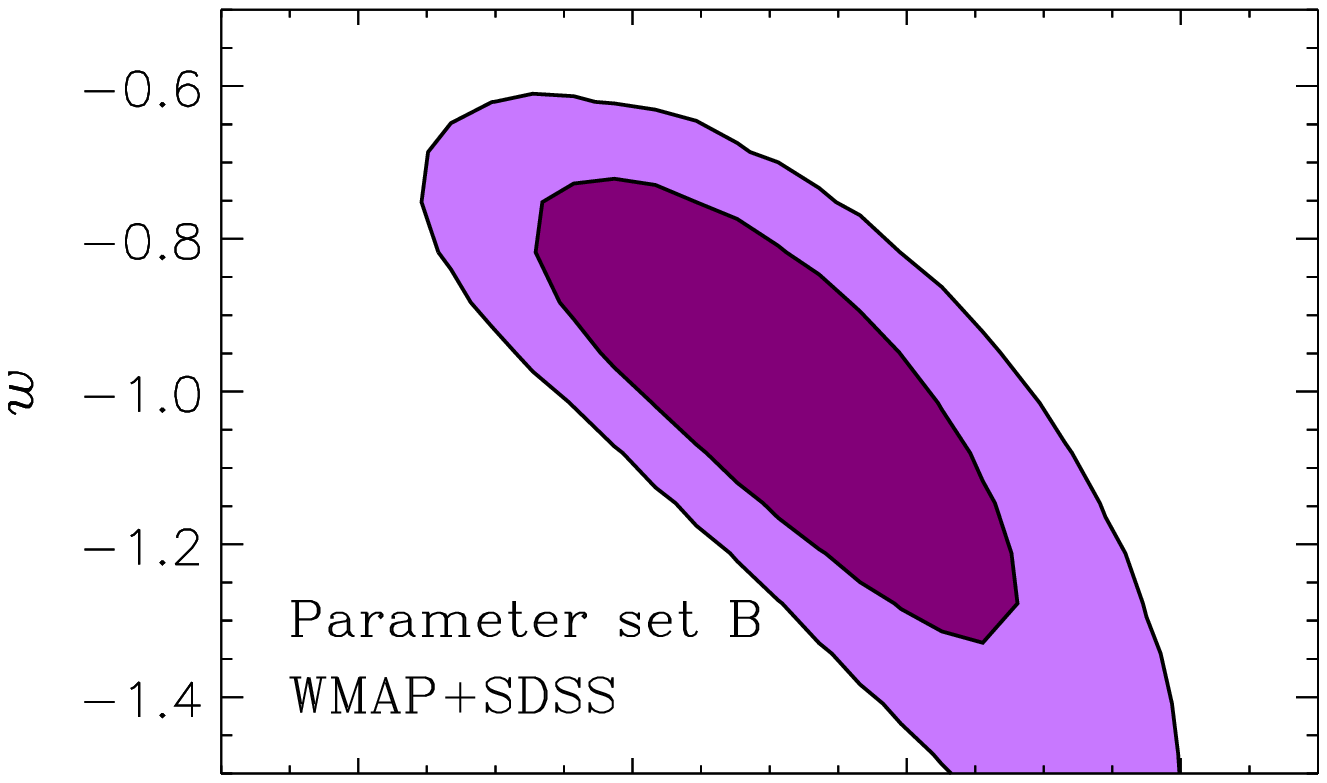,width=8.5cm}\vspace*{-1cm}\\
\epsfig{file=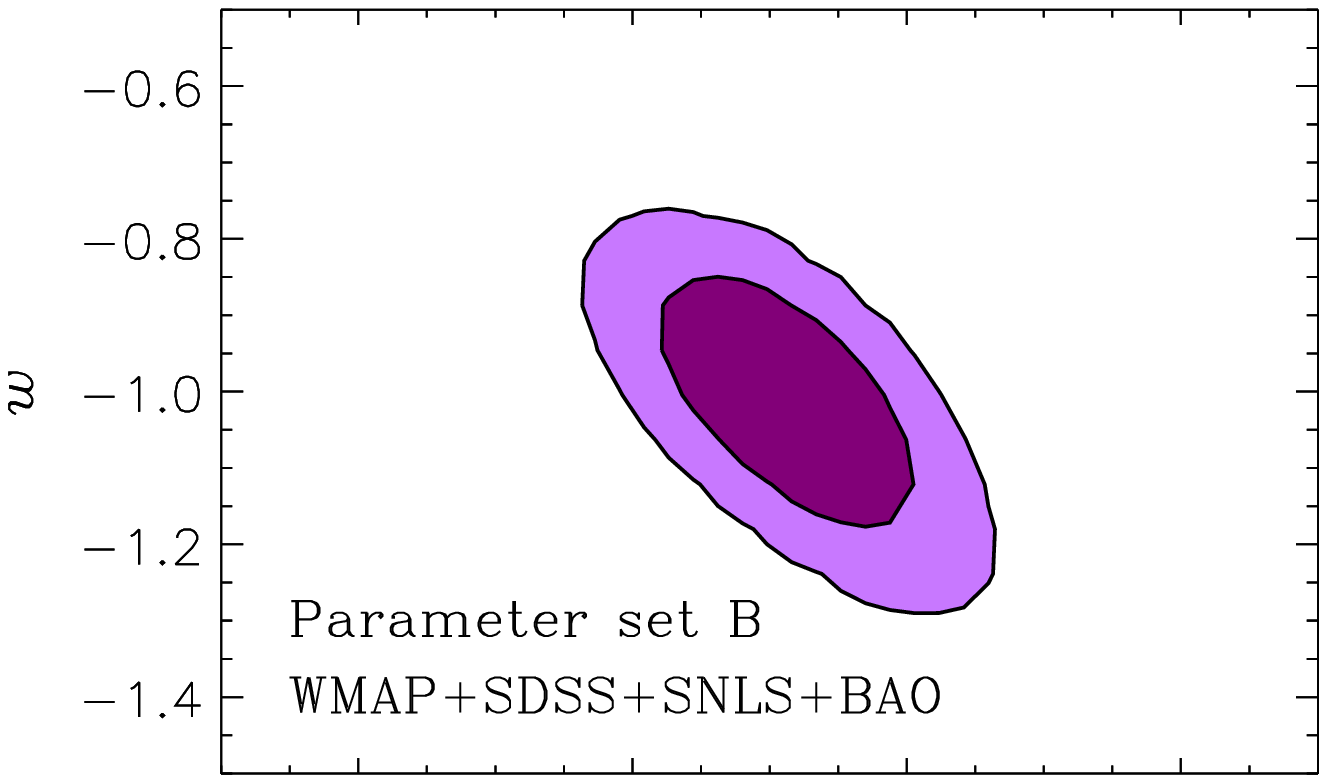,width=8.5cm}\vspace*{-1cm}\\
\epsfig{file=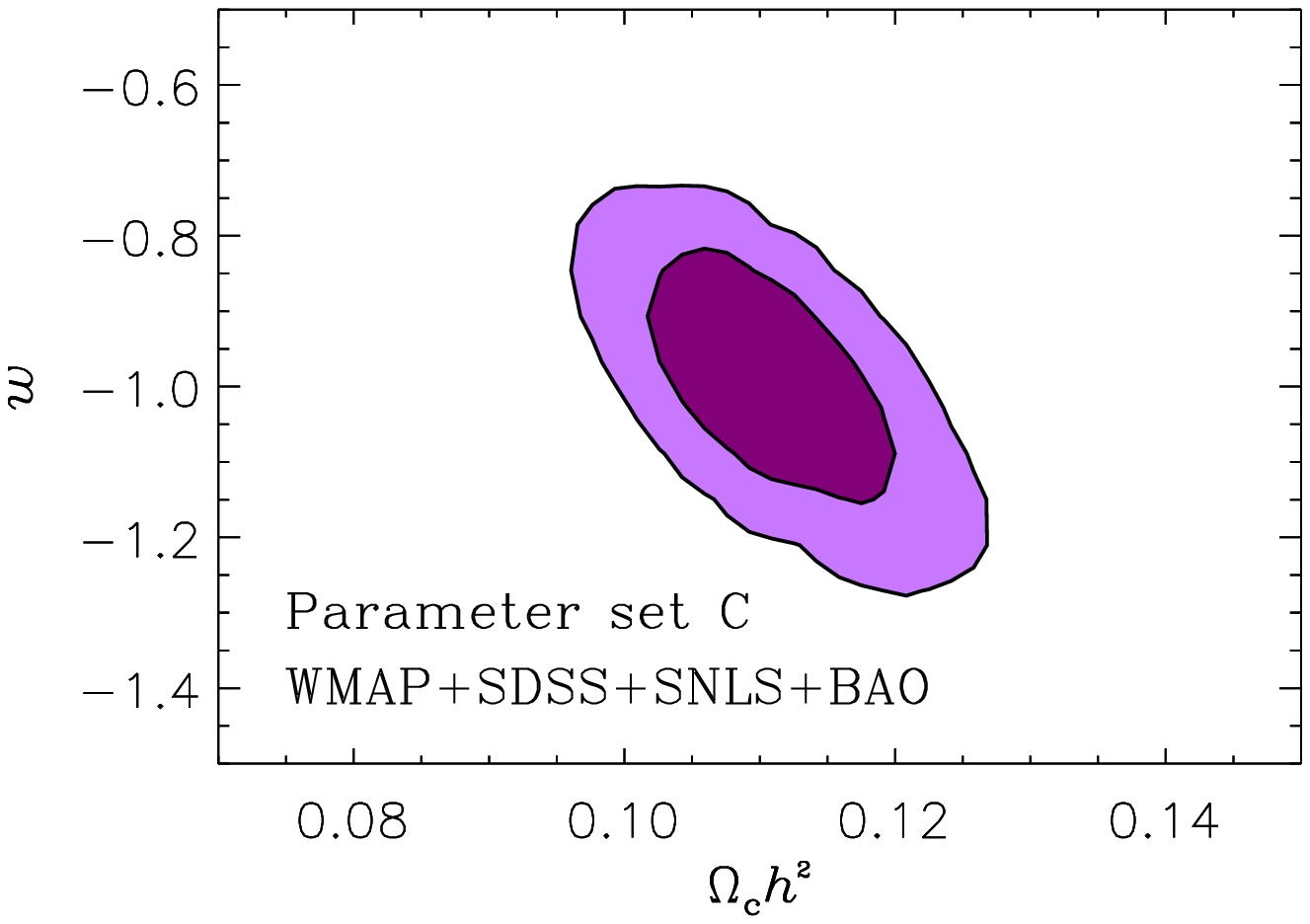,width=8.5cm}
\end{center}
\caption{Two-dimensional marginalised $68\ \%$ and $95 \ \%$ C.L.\
contours for $\Omega_c h^2$ and $w$,
using various parameter and data sets. {\it Top}: Parameter set B, WMAP+SDSS.
{\it Middle}: Parameter set B,
the full data set. {\it Bottom}: Parameter set B with $\alpha_s=0$ (i.e., parameter set C),
the
full data set.}\label{fig:omegac}
\end{figure}

\subsection{Dark matter and dark energy} \label{sec:DMandDE}

In order to derive robust bounds on the physical dark matter and
dark energy properties, all other plausible parameters should be
allowed to vary. With respect to the initial conditions this is
almost impossible since the most general inflationary models
do not necessarily give smooth, power law-like spectra.  Instead,
the primordial power spectrum can have various features
\cite{Starobinsky:1986fx,Adams:1997de,Elgaroy:2003hp,Hunt:2004vt,Covi:2006ci,%
Martin:2003kp,Contaldi:2003zv}
(see also
\cite{Souradeep:1997my,Wang:1998gb,Hannestad:2000pm,Hannestad:2003zs,%
Bridges:2006zm,Shafieloo:2006hs} for more observationally oriented
discussions), which may bias estimates of parameters and their
errors.
Here we present just a small step towards dark
matter and dark energy parameter estimation
 in the context of more general models.

The physical dark matter density $\Omega_c h^2$ is a crucial
input in dark matter model building. A prime example of this is
models with low energy SUSY where the dark matter particle
is usually either
the neutralino or the gravitino. Large regions in parameter space
in these models have been excluded by the fact that the predicted
dark matter density is
too high or too low
\cite{Profumo:2004at,deAustri:2006pe,Ellis:2003si,Battaglia:2003ab,%
Ellis:2003cw,Carena:2006nv,Baltz:2006fm,Aguilar-Saavedra:2005pw,%
Belanger:2005jk,Allanach:2004xn,Steffen:2006hw}.

In the vanilla model
$\Omega_c h^2$ is a very well constrained quantity,
with WMAP+SDSS giving a $68 \ \%$ C.L.\ limit
of $\Omega_c h^2 = 0.1050^{+0.0041}_{-0.0040}$
\cite{Tegmark:2006az}. This corresponds to a relative uncertainty
of $\sigma(\Omega_c h^2)/\Omega_c h^2 \simeq 0.04$.
The SDSS collaboration also provide bounds on $\Omega_c h^2$
in extended models in which one additional parameter is added
to the vanilla parameter set \cite{Tegmark:2006az}. In most cases the
bound on $\Omega_c h^2$ does not change significantly.
However, when either $\Omega_k$ or $w$ is allowed to vary,
$\sigma(\Omega_c h^2)/\Omega_c h^2$ increases to
about $0.06$ \cite{Tegmark:2006az}.

We have taken this investigation further by calculating the bound
on $\Omega_c h^2$ for our various parameter and data sets. In
Fig.~\ref{fig:omegac} we show the joint 2D marginalised constraints
on $\Omega_c h^2$ and $w$ for three different cases using
parameter sets B and C. If only WMAP and SDSS data are used,
a very strong
degeneracy between $\Omega_c h^2$ and $w$ weakens the bounds on
both parameters. This degeneracy is broken when SNIa and BAO are
included (as is also the case with the degeneracy between $f_\nu$
and $w$), yielding strong constraints on both parameters.

\begin{figure}[tb]
\begin{center}
\epsfig{file=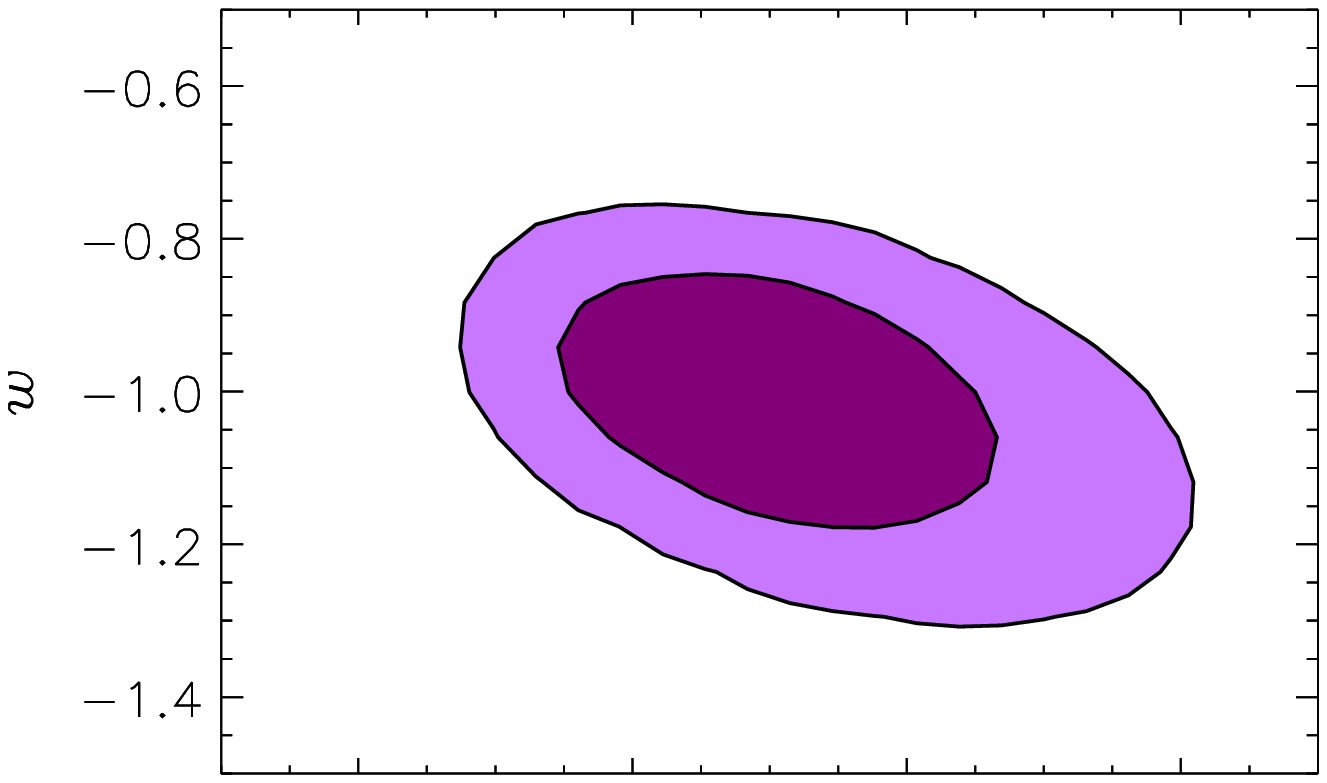,width=8.5cm}\vspace*{-1cm}\\
\epsfig{file=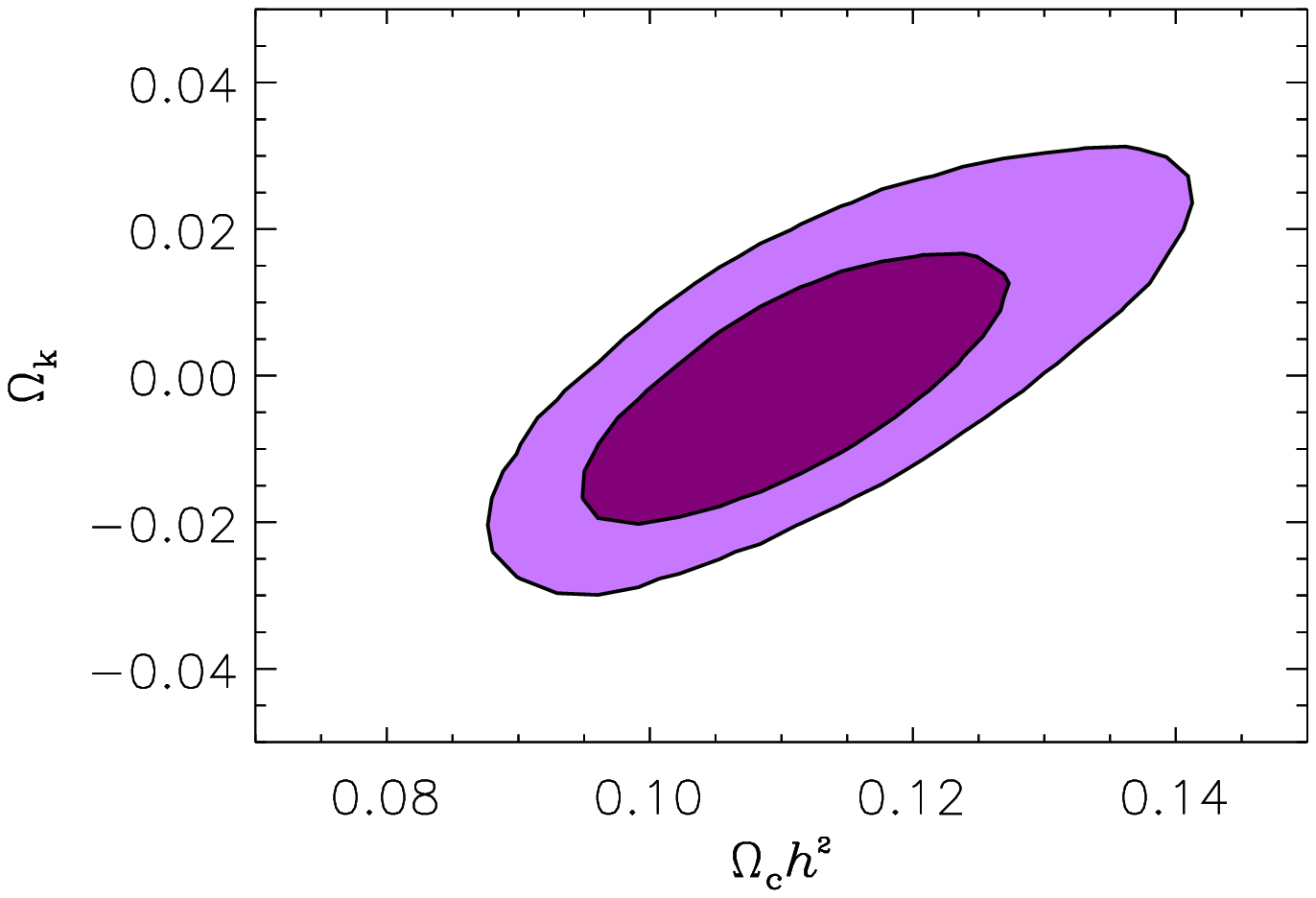,width=8.5cm}
\end{center}
\caption{Two-dimensional marginalised $68\ \%$ and $95 \ \%$ C.L.\
contours for $\Omega_c h^2$, $\Omega_k$, and $w$, using
parameter set A and the full data set.}\label{fig:omegac2}
\end{figure}

When spatial curvature is also allowed to vary,
the bound on $\Omega_c h^2$ does change considerably. Figure
\ref{fig:omegac2} shows the 2D marginalised contours for $\Omega_c
h^2, \Omega_k$ and $w$, using parameter set A and the full data
set. Here we find $\sigma(\Omega_c h^2)/\Omega_c h^2 \simeq
0.1$, so that $0.094 < \Omega_c h^2 < 0.136$,
$-0.022 <\Omega_k < 0.026$, and $-1.19 < w < -0.88$
(1D at $95 \ \%$ C.L.).  It is interesting to compare our more
general constraints on $\Omega_k$
with that given by the SDSS collaboration from a vanilla+$\Omega_k$
fit ($-0.015 < \Omega_k < 0.023$, $95 \ \%$ C.L.) \cite{Tegmark:2006az};
our allowed range is slightly larger even in the light of
additional data from the distance measurements of SNIa and BAO.
We note also that even though the allowed range
for $\Omega_c h^2$ increases considerable with the inclusion of
$\Omega_k$, the same is not true for the dark energy equation
of state parameter $w$. In Table~\ref{tab:omw}, we summarise
the 1D $95 \ \%$ constraints on
$\Omega_c h^2$ and $w$ from Figs.~\ref{fig:omegac} and
\ref{fig:omegac2}

Finally, let us we stress that some caution should be applied
whenever the  dark matter density is used as an input to
constrain models such as the MSSM.
Parameter regions that are excluded in the simplest vanilla model
can easily be allowed in more general models,
even without the introduction of more exotic
features such as isocurvature modes. If one is to take one single
number inferred from cosmological observations as an input to
constrain particle physics models, then the safest approach
is to allow for the possibility that cosmology is not described
by the vanilla model, but by something more general.
From our calculations, we recommend using
$0.094 < \Omega_c h^2 < 0.136$ ($95 \ \%$ C.L.),
but we caution that even this may not be the most conservative
estimate.

\begin{table*}
\caption{\label{tab:omw}The 1D marginalised $95 \ \%$ C.L.\
allowed ranges for $\Omega_c h^2$ and $w$ for various parameter and data
sets.}
\begin{ruledtabular}
\begin{tabular}{lccc}
Parameter set & Data set & $\Omega_c h^2$ & $w$ \\
\hline
B & WMAP+SDSS & $0.092 \to 0.136$ & $-1.44 \to -0.76$ \\
B & WMAP+SDSS+SNLS+BAO & $0.100 \to 0.123$ & $-1.12 \to -0.87$ \\
C & WMAP+SDSS+SNLS+BAO & $0.100 \to 0.123$ & $-1.11 \to -0.86$ \\
A & WMAP+SDSS+SNLS+BAO & $0.094 \to 0.136$ & $-1.19 \to -0.88$ \\
\end{tabular}
\end{ruledtabular}
\end{table*}

\section{Discussion}                             \label{sec:discussion}

We have performed a detailed study of cosmological parameter
estimation in the context of extended models that encompass a larger
model parameter space than the standard,
flat $\Lambda$CDM cosmology.  Using the 6-parameter vanilla model
as a basis, we include as additional parameters only those
that are physically motivated, such as a nonzero neutrino mass.
We consider a 11-parameter model and subsets thereof,
in contrast with the vanilla+1 approach adopted in most
previous analyses which treats one extra parameter at a time.

In this more general framework, we find that in the context of
standard slow-roll inflation, constraints on the dark matter and
dark energy parameters can be substantially altered. If only CMB
and LSS data are used, the larger parameter space introduces new,
strong parameter degeneracies, e.g., between the physical dark
matter density $\Omega_c h^2$ and the dark energy equation of
state $w$. These degeneracies can be broken to a large extent by
adding type Ia supernova and baryon acoustic oscillation data to
the analysis. However, even with this expanded data set, we
find that the bound on the physical dark matter density $\Omega_c
h^2$ relaxes by more than a factor of two compared to the
vanilla model constraint.

In the same spirit, we have studied how bounds on the
inflationary parameters $n_s$, $r$, and $\alpha_s$ are affected
by the introduction of extra parameters in the analysis.
We find that the simplest $\lambda \phi^4$ model of inflation
is still compatible with all present data at the $95 \ \%$ level,
in contrast with other recent analyses
\cite{Kinney:2006qm,Spergel:2006hy,Tegmark:2006az}.
The source of this apparent discrepancy is a strong degeneracy between the
tensor-to-scalar ratio $r$ and the neutrino fraction
$f_\nu$, the latter of which was fixed at zero in the analyses
of \cite{Kinney:2006qm,Spergel:2006hy,Tegmark:2006az}.
Reversing the argument, if $\lambda \phi^4$ is the true model
of inflation, then it strongly favours a sum of
quasi-degenerate neutrino masses between $0.3$ and $0.5$ eV,
a range compatible with present data from laboratory experiments.
This represents a clear example of how neutrino masses well within
laboratory limits can bias conclusions about other, seemingly unrelated
cosmological parameters.

\section*{Acknowledgments}

We acknowledge use of computing resources from the Danish Center
for Scientific Computing (DCSC).


\end{document}